\documentclass[12pt,letterpaper,preprintnumbers]{article}
\usepackage{amsmath,amssymb,epsfig}
\usepackage[pdftex,bookmarks,bookmarksnumbered,linktocpage,pdfstartview=FitH]{hyperref}
\hypersetup{colorlinks,%
citecolor=blue,%
filecolor=blue,%
linkcolor=blue,%
urlcolor=blue,%
pdftex}
\usepackage[all]{hypcap}
\numberwithin{equation}{section}
\usepackage{cite}

\long\def\symbolfootnote[#1]#2{\begingroup%
\def\thefootnote{\fnsymbol{footnote}}\footnote[#1]{#2}\endgroup}

\newcommand{\be}{\begin{equation}}
\newcommand{\bea}{\begin{eqnarray}}
\newcommand{\eea}{\end{eqnarray}}
\newcommand{\ba}{\begin{array}}
\newcommand{\ea}{\end{array}}
\newcommand{\ee}{\end{equation}}

\expandafter\ifx\csname mathbbm\endcsname\relax

\else

\fi \textheight 22cm \textwidth 15cm \topmargin 1mm \oddsidemargin
5mm \evensidemargin 5mm

\begin{document}

\begin{titlepage}
\hfill{BRX-TH 624}

\vspace*{20mm}

\begin{center}
{\Large{\bf Instantaneous Thermalization in Holographic Plasmas}}

\vspace*{16mm}

{Hajar Ebrahim${{}^{1,2}}$\symbolfootnote[1]{hebrahim@physics.harvard.edu} ,
Matthew Headrick${{}^{1}}$\symbolfootnote[2]{headrick@brandeis.edu}} 

\vspace*{1 cm}

{\it{${}^1$Theory Group, Martin Fisher School of Physics, Brandeis University,}\\ \it{MS057, 415 South St., Waltham, MA 02453, USA}\\ \it{${}^2$Center for the Fundamental Laws of Nature, Harvard University,}\\ \it{Cambridge, MA 02138, USA}}

\vspace*{1cm}

\end{center}

\begin{abstract}
Thin-shell AdS-Vaidya spacetimes can be considered as holographic models of the thermalization process in strongly-coupled conformal field theories following a rapid injection of energy from an external source. While the expected thermalization time is the inverse temperature, Bhattacharyya and Minwalla have pointed out that bulk causality implies that expectation values of local field-theory observables actually take on their thermal values immediately following the injection. In this paper we study two-point functions, for which the causality argument does not apply. Specifically, we study the Brownian motion of a ``quark" represented by a string stretching from the boundary to the horizon of an $\textrm {AdS}_3$-Vaidya spacetime. Surprisingly, we find that the two-point function also thermalizes instantly. Since Brownian motion is a $\frac{1}{N}$ effect, our result shows that, at least in certain cases, the rapid thermalization property of holographic plasmas persists beyond leading order in $\frac{1}{N}$.


\end{abstract}
\end{titlepage}

\newpage
\tableofcontents

\section{Introduction}
One of the fundamental problems in physics is to understand the dynamics of a thermal system out of equilibrium. The time-dependence nature of the problem makes it very hard to explore. An interesting dynamical system which has prompted a lot of interest is the quark-gluon plasma (QGP) produced at RHIC experiment \cite{Shuryak:2003xe,Shuryak:2004cy}. Modeling based on near-ideal hydrodynamics suggests very rapid local thermalization after the collision \cite{Heinz:2004pj}. The process of local equilibration from an initial far from equilibrium state needs to be understood. Since QGP is a strongly coupled non-Abelian plasma, AdS/CFT correspondence \cite{Maldacena:1997re,Witten:1998qj,Gubser:1998bc} provides a powerful tool to explore the dynamics of this system. 

In thermal field theories stable stationary solutions that describe the systems in equilibrium reveal the late-time behaviour of the system. In gauge/gravity duality this corresponds to a stationary black hole in the dual geometry. An interesting question is to explain how the system goes back to equilibrium after being slightly perturbed. From field theory point of view this question has been investigated in two regimes that have some overlap in the long wave-length limit. One is linear response regime which explores the system with very small fluctuations compared to the microscopic scale of the system. In this regime the evolution of the system is explained by linear response theory and fluctuation-dissipation theorem \cite{kubo}. This calculation is based on computing the correlation functions in the field theory. AdS/CFT provides us with an efficient tool to obtain the correlation function of the operators in the CFT using the bulk dual. The calculation of the retarded correlation functions in AdS/CFT was first initiated in \cite{Son:2002sd}. More references on this can be found in the very nice review \cite{Hubeny:2010ry}. From the statistical point of view the fluctuation-dissipation theorem has been proved in the AdS/CFT context in the papers by de Boer, Hubeny, Rangamani, Shigemori \cite{deBoer:2008gu} and Son, Teaney  \cite{Son:2009vu}. They have studied the Brownian motion of the quark on the boundary of BTZ black hole using the bulk information. 

The linear response of the system in field theory translates into the behaviour of the linearized fluctuations of the bulk fields on AdS black hole background. The first step in this direction is to study the connection between quasi-normal modes of the black hole and the rate at which the fluctuations return back to equilibrium \cite{Horowitz:1999jd}.  

The other regime which has a more coarse-grained view of the problem is the hydrodynamic or more generally fluid dynamic regime. This describes the long wave length or low energy physics in which the local fluid variables vary very slowly compared to the microscopic scale in a way that preserves the local thermal equilibrium of the system every where. In the gauge/gravity picture this regime has been explored in the interesting works on fluid/gravity duality which was initiated by \cite{Bhattacharyya:2008jc}. 

Now one can ask what happens when the system is far from equilibrium which does not sit in any of the regimes above. It has been argued in the literature that a system out of equilibrium corresponds to black hole formation in the bulk.  Recently this problem has been addressed in interesting work by Chesler and Yaffe \cite{Chesler:2008hg} and also the work by Bhattacharyya and Minwalla  \cite{Bhattacharyya:2009uu}. The former has numerically studied the thermalization in ${\mathcal N}=4$ SYM  by triggering the gravitational wave and perturbing the boundary stress tensor. The latter has analytically explained the formation of the black hole in the bulk using both minimally coupled scalar wave and graviton perturbation which corresponds to the injection of energy on the boundary. They generally argue, based on the bulk causality, that in the process of the black hole or black brane formation the local operators on the boundary thermalize instantly while the non-local operators such as Wilson loops thermalize in longer time which depends on the scale of the loop. We will base our set-up on the result that a rapid injection of energy on the boundary is dual to a thin-shell Vaidya spacetime  \cite{Bhattacharyya:2009uu}. This seems a reasonable toy model to study thermalization on the boundary.  

Motivated by the work  \cite{Bhattacharyya:2009uu}, in this paper we are taking the first step towards addressing thermalization of non-local operators in the holographic field theory where the bulk metric is planar Vaidya spacetime corresponding to the planar black hole formation. An interesting set-up to address this question is to study the Brownian motion of the quark on the boundary, similar to \cite{deBoer:2008gu}. The quark is the end point of a string stretched between the boundary and the horizon. Its fluctuations are described by the Hawking radiation of the black hole \cite{deBoer:2008gu,Son:2009vu} (The basic idea that a string attached to a black hole fluctuates due to the Hawking radiation is discussed in \cite{Lawrence:1993sg , Frolov:2000kx}). As a candidate for non-local operators we will investigate the two point function of the scalar field explaining the fluctuations of the quark. Due to the causal structure of the Vaidya spacetime, retarded Green's function will thermalize instantly. By this we mean that it equals its thermal value when both arguments are taken to be at any time (no matter how soon) after the injection of energy. But one might initially think that the Hadamard function of the field which is the expectation value of the anti-commutator will behave differently. This is due to the fact that in order to calculate the Hadamard function one needs to know the state of the thermal vacuum which encodes the information about the Hawking radiation of the black hole. In general one expects that it takes some time for the Hawking radiation to turn on. 

One of the obstacles that we face is how to calculate the Hawking radiation at finite time where the bulk spacetime is the formation of a black hole. The usual method used in the literature is to compute the Bogoliubov coefficients. In general this is a very hard calculation and has been done only in high frequency or late time limit where the spacetime is stationary. This calculation relies on using geometric optics approximation. But in this paper we would like to address the on-set of Hawking radiation which means calculating a time dependent Hawking radiation that in the late time limit reduces to the constant result for stationary black holes. In this regard an elegant work has been done by Callan, Giddings, Harvey and Strominger (CGHS) \cite{Callan:1992rs} where they use the trace anomaly \cite{Christensen:1977jc} to calculate the time dependent Hawking radiation in the formation of 2-dimensional linear dilaton black hole. This method can not be used in our system due to the lack of diffeomorphism-invariance in the 2-dimensional worldsheet calculation. 

In this paper we introduce a new way to overcome this problem and obtain the Hadamard function of the scalar field in a general time-dependent background. This is a fairly simple method but one that has not been used in the literature before, as far as we are aware. The idea is to calculate the Hadamard function of the fields in the vacuum or the spacetime before the formation of the black hole. And then, using the fact that the Hadamard function solves the equation of motion in both variables, we propagate it from the shock wave to the black hole region using the black hole retarded Green's function. We thus replace the problem of solving the wave equation an infinite number of times, in order to find a complete set of solutions and compute the Bogoliubov coefficients, with the problem of solving the wave equation just once but in two variables, which is much more tractable in a general time-dependent background (especially a simple one such as thin-shell Vaidya). We will introduce and check this method with the known result of CGHS in subsection \ref{sub:initial}. 

The set-up of the paper is the planar BTZ black hole formation in the bulk. We will show, using the method mentioned in the previous paragraph, that the two-point function or more specifically the Hadamard function of the fluctuation modes which is a non-local operator thermalizes instantly after the shock wave (subsection \ref{sub:result}). This is an interesting result which seems to be at odds with the arguments of \cite{Bhattacharyya:2009uu}. A simple observation that makes this result less surprising is the general discussion of a Rindler observer. If an observer who is at rest in flat space and starts to uniformly accelerate, calculates the two-point function of the fields on its worldline he will measure a thermal two-point function.  

We will also see that using the method of propagation, the same result of instantaneous thermalization of the Hadamard function is obtained in the 2-dimensional set-up of ${\textrm{AdS}}_2$ black hole formation. This is not a new result and has been previously observed in an interesting paper by Spradlin and Strominger \cite{Spradlin:1999bn}. This result applies to the transverse fluctuations of the string in unwarped directions, such as the $S^3$ or $M^4$ directions in $AdS_3\times S^3\times M^4$ where $M^4$ can be $T^4$ or $K3$.

To clarify more the result of this paper we mention that if one wants to use holographic models to understand (even qualitatively) the behaviour of the real quark-gluon plasma, such as its apparent rapid thermalization, one has to understand the $\frac{1}{N}$ effect, which are potentially very important because in QCD $N$ is just 3. Whereas previous works studying thermalization in holographic plasmas have considered the classical dynamics of the bulk spacetime, here we are considering a quantum effect, which from the boundary point of view is a $\frac{1}{N}$ effect. While the particular model we study is not a realistic model of the QGP, it is suggestive that, in our toy model, the rapid thermalization survives at the level of $\frac{1}{N}$ effect. 

An immediate question that comes to mind is to study the same set-up but in higher dimensional Vaidya spacetimes to see if one can observe the same result of instantaneous thermalization. Or this result is particular to 3-dimensional setting. It can be asked why after doing a complicated calculation for the Hadamard function in ${\textrm{AdS}}_3$ Vaidya background we get such a simple answer. Is there any way to organize the calculations that makes the behaviour of the Hadamard function clear. One can also look at the Hadamard function of the scalar field on the Vaidya background instead of the 2-dimensional worldsheet calculation. This can be done in the 3-dimensional set-up of this paper or other higher dimensional black holes. One may also ask what happens to the bulk Hadamard function after the shock wave. In the ${\textrm{AdS}}_2$ black hole formation the bulk Hadamard function thermalizes instantly too. But we have not done this calculation in the BTZ formation set-up. Another interesting approach to thermalization is to study other non-local operators such as Wilson loops. From field theory point of view one might also ask if this observation only applies to holographic strongly coupled plasmas or in any strongly coupled field theory this rapid thermalization happens. All of these paths are very interesting to investigate.  

 \section{The set-up of the problem}\label{sec:set-up}
In this section we will explain the set-up of the problem. We would like to study the Brownian motion of a quark on the boundary where the bulk dual is the 3-dimensional planar Vaidya background. Regarding the result of \cite{Bhattacharyya:2009uu} this can be a good toy model to study thermalization of non-local operators.  The quark on the boundary is in fact the end point of a string which is stretched between the boundary and the horizon. Therefore the fluctuations of the quark are described by the fluctuations of the transverse modes of the string. These fluctuations originate from the Hawking radiation of the black hole  which propagates along the string  \cite{deBoer:2008gu,Son:2009vu}. In fact it has been shown that the modes on the string worldsheet which attaches to the horizon are thermally excited by the black body radiation of the black hole  \cite{Lawrence:1993sg , Frolov:2000kx}.  In this section we will closely follow the notation of  \cite{deBoer:2008gu}.     

The background metric can be written as 
\be
\label{original metric}
ds^2 = g_{\mu\nu}(x) dx^\mu dx^\nu + G(x) dX^2~,
\ee   
where $x^\mu = t,r$. We would like to work in the static gauge in which the worldsheet coordinates are identified with the spacetime coordinates. Therefore one can assume that the string is streched along the $r$ direction and its small fluctuations are in the transverse direction $X$. We consider $g_s$, the string coupling, to be very small and work in the limit where the motion of string is described in the probe approximation which means the backreaction of the string on the background geometry can be ignored. Thus the string $\sigma$-model action in the absence of the B-field can be expanded to quadratic order in $X$ as
\bea
\label{action}
S_{NG} &=& - \frac{1}{2\pi \alpha'} ~\int d^2x~ \sqrt{- det \gamma_{\mu\nu}}~,\\
&\approx&  - \frac{1}{4\pi \alpha'} ~\int d^2x~ \sqrt{- g} g^{\mu\nu} G \partial_\mu X \partial_\nu X~,\nonumber
\eea  
where $\gamma_{\mu\nu}$ is the induced metric on the worldsheet. With this set-up the Brownian motion of a quark on the boundary can be addressed by doing a 2-dimensional worldsheet calculation with a Vaidya metric.  

The bulk spacetime is described by a 3-dim Vaidya metric which can be written as
\be
\label{vaidya metric}
ds^2 = - (\frac{r^2}{l^2} - M \Theta(v)) dv^2 + 2 dv dr + \frac{r^2}{l^2} dX^2~.
\ee
where $M(v) = M \Theta(v)$ is a step function
\bea
M(v) &=& 0 , ~~~~~~v<0~,\cr
M(v) &=& M , ~~~~~~v>0~, 
\eea
and M is the mass of the black hole after formation. The metric has been written in ingoing Eddington-Finkelstein coordinate, $(v,r)$. The advantage of writing the metric in this coordinate system is that it is non-singular on the horizon and the coordinates $(v,r)$are well-defined all over the space. $v=t+r^*$ is the ingoing null coordinate which ranges over $(-\infty,\infty)$ and $0< r < \infty$. $r^*$ is called the tortoise coordinate and is defined as
\be
\frac{dr^*}{dr} = \sqrt{\frac{-g_{rr}}{g_{tt}}}~.
\ee
Therefore we have
\bea
{\textrm{AdS}}_3&:&~~~~~~r^*= - \frac{l^2}{r}~,\cr
\cr
{\textrm{BTZ}}&:&~~~~~~r^*=\frac{l^2}{2 r_H} ~{\textrm{ln}} (\frac{r-r_H}{r+r_H})~,
\eea
where $r_H = \sqrt{M} l$ is the black hole horizon radius and $-\infty <r^* < 0$.
\begin{figure}
\centering
\includegraphics[width=9 cm]{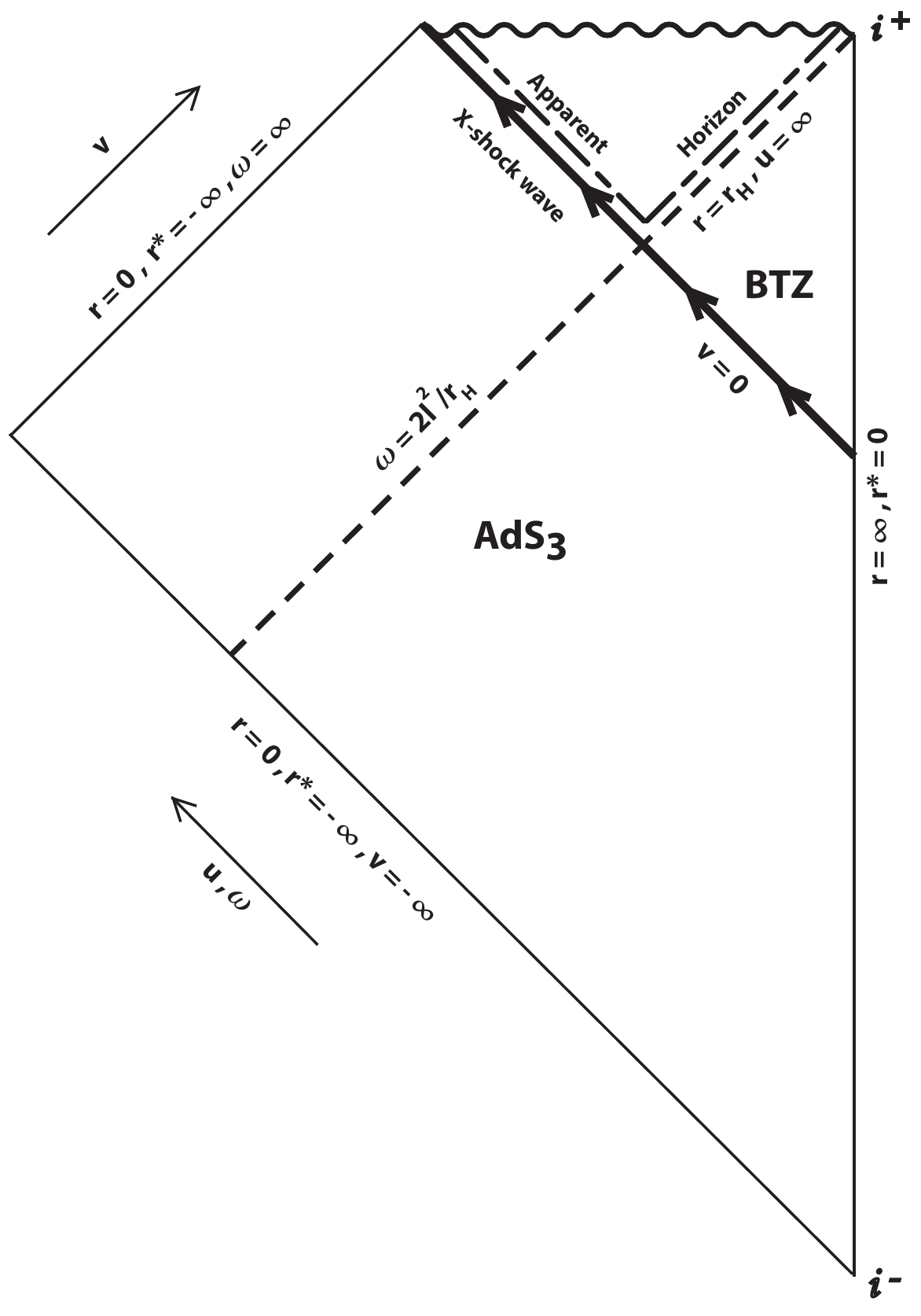}
\caption{\small{Penrose diagram of the BTZ black hole formation. The dashed line is the event horizon and the arrow shows the shock wave which denotes the collapse of null matter.}}
\label{fig:penrose}
\end{figure} 
  
The metric (\ref{vaidya metric}) explains the formation of a black hole with mass M by the collapse of null matter (shock wave) at $v=0$. The shock wave is stretched in the $X$ direction. One can see due to the $r^2$ term in the third component of the metric, the shell shrinks as $r\rightarrow 0$. The bulk spacetime is ${\textrm{AdS}}_3$ before $v=0$ and non-rotating planar BTZ black hole after it. By planar BTZ black hole we mean  $X\in {\mathbb{R}}$, instead of normal definition for BTZ which is $X=l \varphi$ with $\varphi=\varphi+2 \pi$. 

Basically we would like to work with the null coordinates $(v,u)$ where $u = t- r^*$ is the outgoing null coordinate and ranges over $-\infty$ to $\infty$. But it is not continuous across the shock wave and we have to find the relation between outgoing null coordinate in ${\textrm{AdS}}_3$ and BTZ. Let's define $u$ as a function of $u(v,r)$. Regarding the fact that we want to write the metric in null coordinates, we have $ds^2 =0$ when $du=0$. Therefore in order to calculate $r$ as a function of $v$ and $u$ we have to solve the differential equation
\be
2\frac{\partial r}{\partial v} -\frac{1}{l^2}(r^2 - r_H^2 \Theta(v)) =0~. 
\ee  
This equation can be solved for the two regions before and after the shock wave and we get 
\bea
r &=& \frac{-2 l^2}{v+f_1(u)}~;~~~~~~v<0~,\cr
 \cr
 r &=& -r_H~\frac{1+e^{-\frac{r_H}{l^2}(v+f_2(u))}}{1-e^{-\frac{r_H}{l^2}(v+f_2(u))}}~;~~~~~~v>0~.
\eea
Since $r$ is the well-defined coordinate in all the space-time so it should be continuous across the shock wave at $v=0$. Imposing this condition on the above result we get
\be
\label{u-trans}
f_2(u) = \frac{-l^2}{r_H}  {\ln} (\frac{1-\frac{r_H}{2 l^2} f_1(u)}{1+\frac{r_H}{2 l^2} f_1(u)})~.
\ee 
\noindent We have the freedom to choose $f_1(u)$ and $f_2(u)$ in terms of outgoing null coordinate in ${\textrm{AdS}}_3$ and BTZ spacetimes. If we choose $f_1(u)=-\omega$ and $f_2(u)=-u$ where $\omega$ and $u$ are the outgoing null coordinate in ${\textrm{AdS}}_3$ and BTZ, respectively, we get
\be
\label{u-relation}
\omega= \frac{2 l^2}{r_H} ~{\tanh} (\frac{r_H}{2 l^2} u)~.
\ee
Therefore at the horizon of the black hole which corresponds to $u=\infty$ we have $\omega = \frac{2 l^2}{r_H}$.  

The Penrose diagram of the spacetime is shown in figure (\ref{fig:penrose}). The dashed line is event horizon and dash-dotted line is the apparent horizon. The role of the apparent horizon in non-equilibrium systems has been mentioned in \cite{Chesler:2008hg}.

In the following subsections we explain the mode expansions in each space, separately.   

\subsection{String Mode Expansion in ${\textrm{AdS}}_3$}   
\label{subsec:Mode Expansion in $AdS_3$}

Since we would like to study the behaviour of the correlation function of the fluctuations of the quark on the boundary we need to obtain the mode expansion of the transverse mode of the string in the two regions of the Vaidya spacetime. We start with ${\textrm{AdS}}_3$ metric in the Poincar\'e patch which is
\be
\label{AdS3metric}
ds^2 = - \frac{r^2}{l^2} dt^2 + \frac{l^2}{r^2} dr^2 + \frac{r^2}{l^2} dX^2~.
\ee
The string is stretched along $r$ direction and the worldsheet coordinates are assumed to be  $(t,r)$ (static gauge). The 2-dimensional worldsheet  equation of motion is  
\be
\label{eom}
\frac{1}{\sqrt{-g}} \partial_\mu \bigg{(}\sqrt{-g} ~G ~g^{\mu\nu} ~\partial_\nu X\bigg{)}=0~,
\ee  
where compared to (\ref{original metric}) we have 
\be 
G(x) dX^2 = \frac{r^2}{l^2} dX^2~.
\ee
Note that in a higher dimensional spacetime the equation of motion could be written as 
\be
\frac{1}{\sqrt{-g}} \partial_\mu \bigg{(}\sqrt{-g} ~G_{IJ} ~g^{\mu\nu} ~\partial_\nu X^J\bigg{)}=0~.
\ee
Using (\ref{AdS3metric}) the equation of motion reduces to 
\be
- \partial^2_t  X + {r^*}^2 \partial_{r^*} (\frac{1}{{r^*}^2} \partial_{r^*} X)=0~.
\ee
This equation can be even more simplified if we assume 
\be
\label{eq:phi}
X(t,r) = \frac{\phi (t,r)}{r}~ .
\ee
Therefore we get 
\be
- \partial^2_t  \phi + \partial^2_{r^*} \phi - \frac{2}{{r^*}^2} \phi=0~ .
\ee
One can see that this equation is similar to the Schr\"odinger equation with the potential
\be
V(r^*) = \frac{-2}{{r^*}^2} = \frac{- 2 r^2}{l^4}~ .
\ee
It is infinite at $r\rightarrow \infty$ which is due to the fact that the space-time is asymptotically ${\textrm{AdS}}_3$. 

Let's assume that the function $X(t,r)$ is separable. Due to the fact that the metric is time-independent we can write the classical solution to the wave equation for the string modes in the form
\be
u_{\omega}(t,r) = e^{ - i \omega t} f_{\omega}(r)~ .
\ee
where $\omega>0$.
Therefore the string modes are given by the function
\be
u_\omega (t,r) = A \bigg{[}\frac{(r + i\omega l^2)}{r} e^{ i \omega r^*}
 + B \frac{(r - i \omega l^2)}{r} e^{ - i \omega r^*}\bigg{]} e^{-i \omega t}~ , 
 \ee
where $A$ and $B$ are constants and should be fixed by the boundary conditions. In order to have a finite mass for the external particle on the boundary in AdS/CFT context, we need to impose a UV cut-off near the boundary. We choose it as $r_c >>1$. In order to fix the constant $B$ in the wave equation we impose the Neumann boundary condition at the cut-off
\be
\partial_r u_\omega(r,t)|_{r_c} =0 ~.
\ee
Therefore we get
\be
B = - e^{2 i \omega r^*_c}~ ,
\ee  
which is a pure phase. 

The next step is to normalize the modes in order to fix the free constant $A$. This can be done by imposing the normalization conditions 
\be
(u_\omega,u_{\omega'}) = - (u^*_\omega,u^*_{\omega'}) = \delta(\omega-\omega')~;~~~~~~~(u_\omega,u^*_{\omega'}) = (u^*_\omega,u_{\omega'})=0 ~.
\ee
The definition of the inner product is given by
\be
(f,g)_{\Sigma} = - \frac{i}{2 \pi \alpha'}~ \int_{\Sigma} dx \sqrt{h} n^{\mu} G_{IJ} (f^I \partial_\mu g^{J *}-\partial_\mu f^I g^{J *}) ~,
\ee
where $\Sigma$ is a Cauchy surface, $n^\mu$ is the future pointing unit normal vector to  $\Sigma$ and $h_{ij} $ is the induced metric on $\Sigma$. The Cauchy surface is given by $S = t-t_0$ which is  a constant $t$ hypersurface. Therefore the definitions for the normalization condition reduces to
\be
\label{norm}
(u_{\omega},u_{\omega'}) =  - \frac{i}{2 \pi \alpha'}~ \int_{\Sigma} dr ~\frac{r^2}{l^2} g_{rr} (u_{\omega} \partial_t u_{\omega'}^* - u_{\omega'}^* \partial_t u_{\omega})~ ,
\ee
which leads to  
\be
A = \frac{1}{\omega l}\sqrt{\frac{\alpha' }{2\omega}}~.
\ee
For the future reference we mention that in case we consider the modes to be slightly massive ($\mu \rightarrow 0$) we get 
\be
A = \frac{1}{k l}\sqrt{\frac{\alpha' }{2 \omega}}~,
\ee
where $\omega=\sqrt{k^2+\mu^2}$. Therefore the mode expansion is given by
\be
\label{ads-mode}
u_k (t,r) = \frac{1}{k l}\sqrt{\frac{2 \alpha' }{\omega}} e^{i k r^*_c} \bigg{[}{\sin}k(r^*-r^*_c) - k r^*  ~{\cos} k(r^*-r^*_c) \bigg{]}e^{-i\omega t} ~.
\ee 
In the limit where we set $r^*_c\rightarrow 0$ which means we move the boundary to infinity the solution reduces to
\be
u_k (t,r)  = \frac{1}{k l}\sqrt{\frac{2 \alpha' }{ \omega}} \bigg{[}{\sin} k r^* - k r^* ~{\cos} k r^* \bigg{]}e^{-i\omega t}~ .
\ee

Now that we have solved the classical wave equation for the transverse fluctuations of the string in ${\textrm{AdS}}_3$ background, we can write the quantum field for the scalar modes as 
\be
\label{ads-field}
X(r,t) = \int_0^\infty dk~ [a_k u_k (t,r) + a_k^\dagger u^*_k (t,r)]~ ,
\ee
in terms of annihilation and creation operators which satisfy 
\be
\label{operators}
[a_k,a_{k'}] = [a^\dagger_k,a^\dagger_{k'}] =0~;~~~~[a_k,a^\dagger_{k'}] = \delta (k-k') ~.
\ee

\subsection{String Mode Expansion in BTZ }
\label{subsec:Mode Expansion in BTZ}
The same calculation can be done for the BTZ background
\be
ds^2 = - \frac{r^2-r_H^2}{l^2} dt^2 + \frac{l^2}{r^2-r_H^2} dr^2 + \frac{r^2}{l^2} dX^2~ .
\ee
The equation of motion for the string fluctuations gets the form
\be
-\partial_t^2 X + \frac{1}{ r^2}~ \partial_{r^*} (r^2  \partial_{r^*} X)=0 ~.
\ee
which gets more simplified to 
\be
-\partial_t^2 \phi + \partial_{r^*}^2 \phi - 2 \frac{ r_H^2}{l^4}~\frac{1}{\sinh^2 (\frac{r_H r^*}{l^2})} \phi =0 ~.
\ee
where $\phi$ is defined in (\ref{eq:phi}).
Note that similar to the ${\textrm{AdS}}_3$ case we can read the potential which is given by the third term in the above equation. It vanishes at the horizon and diverges at infinity, as we expect. 

The solution to the equation of motion is the sum over the ingoing and outgoing modes  \cite{deBoer:2008gu}
\be
u_\omega = A\bigg{[}\frac{r_H}{r_H+ i\omega l^2} \frac{r+ i \omega l^2}{r} \bigg{(}\frac{r-r_H}{r+r_H}\bigg{)}^{\frac{ i \omega l^2}{2 r_H}} + B \frac{r_H}{r_H- i\omega l^2} \frac{r- i \omega l^2}{r} \bigg{(}\frac{r-r_H}{r+r_H}\bigg{)}^{\frac{- i \omega l^2}{2 r_H}}\bigg{]} e^{-i\omega t} ~.
\ee
Similar to what we did in the ${\textrm{AdS}}_3$ case, we impose the Neumann boundary condition at the UV cut-off and we get  
\be
B =  \frac{r_H - i\omega l^2}{r_H + i\omega l^2} ~\frac{r_H^2 + i\omega l^2 r_c}{r_H^2 - i\omega l^2 r_c}~e^{2i\omega r^*_c} ~.
\ee
We also use the normalization condition to fix the coefficient $A$ in the mode expansion. After some  calculations we get
\be
A = \frac{l}{r_H}\sqrt{\frac{\alpha'}{2 \omega}}~.
\ee

In the future calculations we might need to regularize the theory by considering the modes to be slightly massive ($\mu\rightarrow 0$). Therefore the mode expansion gets the form 
\bea
\label{btz-mode}
u_k&=& l \sqrt{\frac{2 \alpha'}{\omega}}~\frac{e^{i k r^*_c}}{r (r_H+i k l^2)(r_H^2-i k l^2 r_c)}\\
~~\cr
&\bigg{[}&(r_H^2 r+ k^2 l^4 r_c) ~{\cos}~k (r^*-r^*_c) -k l^2(r_H^2 - r_C r) ~{\sin}~k (r^*-r^*_c)\bigg{]} e^{-i\omega t}~ .\nonumber
\eea
In the limit where we set $r^*_c\rightarrow 0$ which means we move the boundary to infinity the solution reduces to
\be
\label{inflimit}
u_k= l \sqrt{\frac{2 \alpha'}{\omega}}~\frac{1}{r (r_H+i k l^2)} ( k l^2 ~{\cos} k r^* + r ~{\sin} k r^*) e^{-i\omega t} ~.
\ee

Similar to the ${\textrm{AdS}}_3$ case the quantum field for the scalar modes we obtained classically is
\be
X(r,t) = \int_0^\infty dk~ [b_k u_k (t,r) + b_k^\dagger u^*_k (t,r)] ~.
\ee
Note that the creation and annihilation operators are not the same as the ones in ${\textrm{AdS}}_3$ space due to the formation of the black hole and the different vacuum state. But they satisfy the same relations as (\ref{operators}).

\section{Green's Function}  
To explain the Brownian motion of the quark on the boundary we need to calculate the two-point function  which describes the correlation between the two points. Our background is a Vaidya spacetime and due to its time dependence, this calculation can not be done explicitly. In the following section we will introduce a way to do this. But before that we need to obtain the Green's function and Hadamard function of the two spaces, ${\textrm{AdS}}_3$ and BTZ, individually. The commutator (and the Green's function which is obtained from the commutator of the fluctuation modes) does not depend on the state, because in a free theory it is a $c$-number. This can be easily seen from equation (\ref{com-ref}).  But the Hadamard function which is defined as the expectation value of the anti-commutator of the fluctuation modes, (\ref{anti}),  depends on the state of the vacuum whether it is thermal. In the ${\textrm{AdS}}_3$ case this is not a problem since  the vacuum is not thermal. But in the BTZ background we should use the fact that the vacuum is thermal and as we will see the calculation relies on the black body radiation of the black hole. The Hadamard function of the modes in BTZ background has been obtained in the limit $r^*_c\rightarrow 0$. These calculations will be used later to get the Hadamard function of the modes on the boundary in the Vaidya background. 

\subsection{${\textrm{AdS}}_3$ Two-Point Functions} 
In the previous section we studied the mode expansion in ${\textrm{AdS}}_3$ space, equation (\ref{ads-field}). We can use it to obtain the Green's function (commutator of the modes which is also called Pauli-Jordan or Schwinger function) and the Hadamard function of the modes. 

\subsubsection{Green's Function}

\begin{figure}
\centering
\includegraphics[width=15 cm]{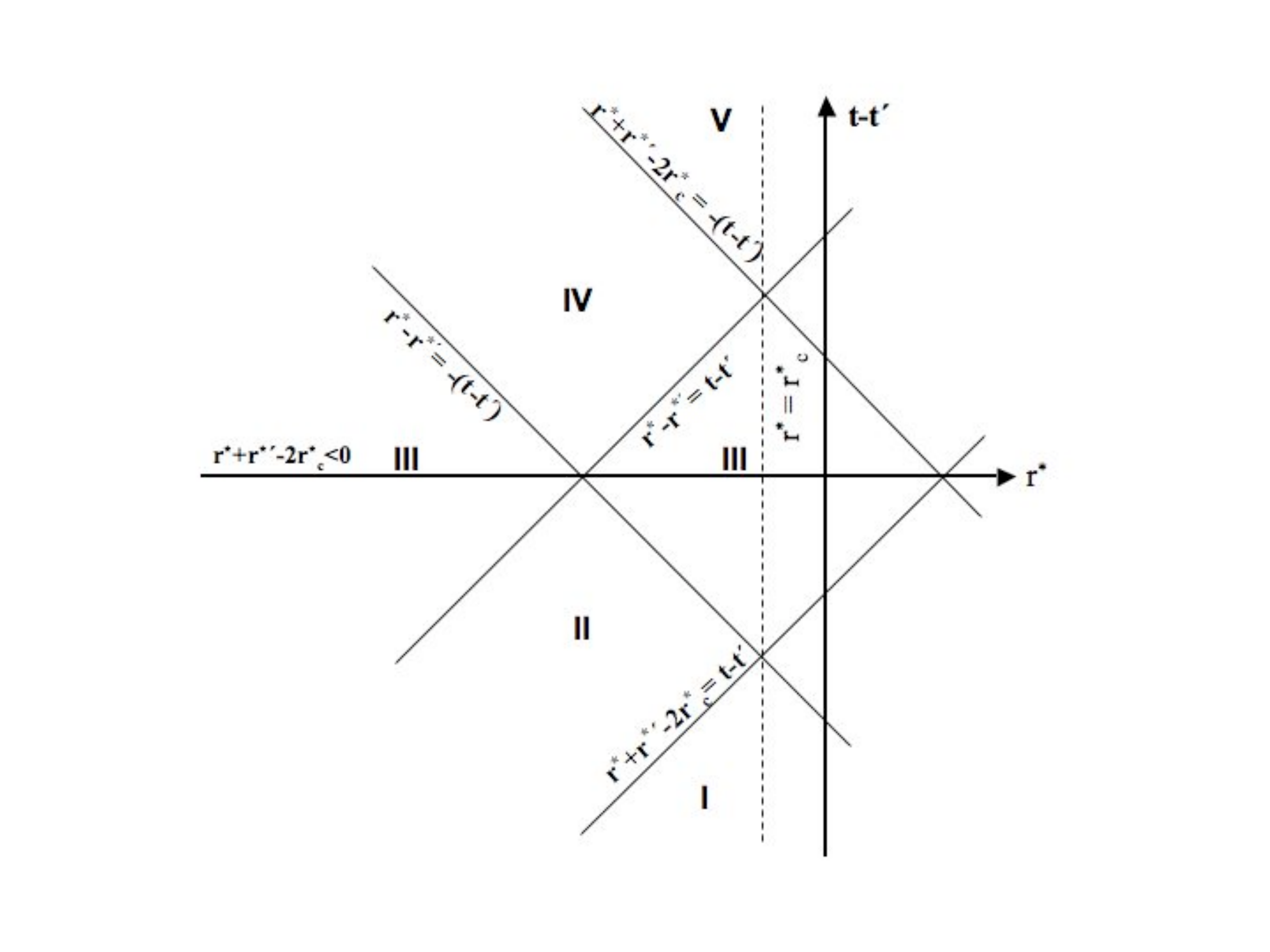}
\caption{\small {Different regions of the light cone. The dashed line is $r^* = r^*_c$ }}
\label{fig:green}
\end{figure}

In general the commutator of the modes is not state-dependent and for the mode expansion (\ref{ads-field}) is given by
\be
\label{com-ref}
\langle 0 | [X(t,r) , X(t',r')] | 0 \rangle = 2 i~ \int_0^\infty dk~ {\textrm{Im}} (u_k (t,r) u^*_k (t',r')) ~.
\ee
Using the solution to the wave equation given in (\ref{ads-mode}) one can see that  the integrals are all finite and we don't need to regularize them. So we consider $\mu=0$ and therefore $k=\omega$ in (\ref{ads-mode}). We can divide the causal structure of our space into different regions which are shown in figure (\ref{fig:green}). The regions are explained by
\bea
I&:&~~r^*+{r^*}'-2 r^*_c > \Delta t ~,\cr
\cr
II&:&~~|r^*-{r^*}'| > \Delta t > r^*+{r^*}'-2 r^*_c~ , \cr
\cr
III&:&~~r^*-{r^*}' > |\Delta t|~;~~~r^*+{r^*}'-2 r^*_c<- |\Delta t|~ ,\cr
\cr
IV&:&~~|r^*-{r^*}'| > -\Delta t > r^*+{r^*}'-2 r^*_c~ ,\cr
\cr
V&:&~~r^*+{r^*}'-2 r^*_c > -\Delta t ~.
\eea
Therefore the commutators in different regions of the figure (\ref{fig:green}) are given by
\bea
I&:&~~~\langle 0 | [X(t,r) , X(t',r')] | 0 \rangle = \frac{ 2 i \alpha' \pi}{l^2} {r^*_c}^2 ~,\cr
\cr
II&:&~~~\langle 0 | [X(t,r) , X(t',r')] | 0 \rangle = \frac{ i \alpha' \pi}{2 l^2} ({r^*}^2 + {{r^*}'}^2 -\Delta t (4 r^*_c + \Delta t)) ~,\cr
\cr
III&:&~~~\langle 0 | [X(t,r) , X(t',r')] | 0 \rangle = \frac{- 2 i \alpha' \pi}{l^2} r^*_c \Delta t ~,\cr
\cr
IV&:&~~~\langle 0 | [X(t,r) , X(t',r')] | 0 \rangle = \frac{- i \alpha' \pi}{2 l^2} ({r^*}^2 + {{r^*}'}^2 -\Delta t (-4 r^*_c + \Delta t))~ ,\cr
\cr
V&:&~~~\langle 0 | [X(t,r) , X(t',r')] | 0 \rangle =\frac{- 2 i \alpha' \pi}{l^2} {r^*_c}^2 ~,
\eea
where we have defined $\Delta t = t-t'$. Causality forces us to have zero commutator (Green's function)  in region III, outside the light cone. But the above result is not zero in that region and it shows that we have to consider the contribution of the zero mode. The zero mode means to consider the mode where the momentum, $k$, is zero. Therefore we have
\be
\label{zeromode}
u_0 =  \frac{1}{l} \sqrt{\frac{\pi \alpha' (-r^*_c)}{\mu}}~ e^{-i\mu t}~ .
\ee
We can write the field for zero mode in terms of creation and annihilation operators as 
\be
X_0(t,t') = u_0 a_0 + u_0^* a_0^\dagger ~,
\ee
where 
\be
[a_0,a_0] = [a_0^\dagger,a_0^\dagger] =0~;~~~~[a_0,a_0^\dagger] =1~;~~~~[a_0,a_{k\neq 0}^\dagger] =0~ .
\ee
Now we can calculate the commutator for the zero mode which is given by
\be
\langle 0 | [X_0(t) , X_0(t')] | 0 \rangle = \frac{2i}{l^2} \alpha' \pi r^*_c ~\frac{{\sin} \mu \Delta t}{\mu} ~.
\ee
In the limit where $\mu\rightarrow 0$ we get
\be
\langle 0 | [X_0(t) , X_0(t')] | 0 \rangle  = \frac{2 i}{l^2} \alpha' \pi r^*_c \Delta t ~.
\ee
Including the zero mode commutator the Green's function is given by
\be
G(r,t;r',t') =  -i \bigg{(}\langle 0 | [X(t,r) , X(t',r')] | 0 \rangle + \langle 0 | [X_0(t) , X_0(t')] | 0 \rangle \bigg{)} ~.
\ee
Thus we have
\bea
I&:&~~~G(r,t;r',t')  = \frac{ 2 \alpha' \pi}{l^2} {r^*_c} (r^*_c + \Delta t)~ ,\cr
\cr
II&:&~~~G(r,t;r',t')  = \frac{\alpha' \pi}{2 l^2} ({r^*}^2 + {{r^*}'}^2 - \Delta t^2 )~ ,\cr
\cr
III&:&~~~G(r,t;r',t')  =0~ ,\cr
\cr
IV&:&~~~G(r,t;r',t')  = \frac{ - \alpha' \pi}{2 l^2} ({r^*}^2 + {{r^*}'}^2 -\Delta t^2 )` ,\cr
\cr
V&:&~~~G(r,t;r',t')  =\frac{- 2  \alpha' \pi}{l^2} {r^*_c} (r^*_c -\Delta t)~.
\eea
We can see that the Green's function satisfies the equation of motion in both variables. We can also see that the retarded, $t-t'>0$, and advanced, $t-t'<0$, Green's functions appear with the opposite signs. Another observation is that the Green's functions in regions II and IV don't depend on the cut-off, $r_c$, due to the contribution of the zero mode. We can also see that the Green's functions at regions I and V are zero when we set $r_c\rightarrow \infty$ or $r^*_c \rightarrow 0$.

\subsubsection{Hadamard Function}
The correlation function between two points in the space can also be explained using  the Hadamard function of the field. For the mode expansion of the fluctuations of the string in ${\textrm{AdS}}_3$ space, eq (\ref{ads-field}), the Hadamard function is given by
\bea
\label{anti}
&&\langle 0 | \{X(t,r), X(t',r')\} | 0 \rangle = 2 \int_0^\infty dk~ \bigg{[}{\textrm{Re}} (u_k (t,r) u^*_k (t',r')) \\
\cr
&+&\int_0^\infty dk' \bigg{(}u_k (t,r) u^*_{k'} (t',r') \langle 0| a^\dagger_{k'} a_k |0 \rangle +u^*_k (t,r) u_{k'} (t',r') \langle 0| a^\dagger_{k} a_{k'} |0 \rangle \bigg{)}\bigg{]}~ .\nonumber
\eea
As it's clear from its definition, for a free theory, the Hadamard function obeys the wave equation in both of its arguments (section 2.7  of \cite{BD}). According to (\ref{anti}) the Hadamard function depends on the state of the vacuum. ${\textrm{AdS}}_3$ is not thermal and we have 
\be
a_k |0\rangle =0 .
\ee
So we get
\be
\langle 0 | \{X(t,r), X(t',r')\} | 0 \rangle = 2 \int_0^\infty dk~ {\textrm{Re}} (u_k (t,r) u^*_k (t',r'))~ .  
\ee
Using the solution to the wave equation in ${\textrm{AdS}}_3$ space, eq (\ref{ads-mode}), 
one can easily see that if we consider massless modes where $\mu=0$, these integrals are divergent in the limit $k\rightarrow 0$. Therefore we should regularize the integrals. The logarithmic divergence in the integrals appears when $k\rightarrow 0$. Therefore we change the lower limit of the integral to $\mu\neq 0$ instead of zero. After the integration we can take the limit $\mu\rightarrow 0$. The final result for the Hadamard function is 
\bea
\langle 0 | \{X(t,r), X(t',r')\} | 0 \rangle &=& \frac{ \alpha'}{2 l^2} \bigg{(}-8 {r^*_c}^2  ~({\ln} \mu ~+\gamma) +12 {r^*_c}^2 -4 r^*_c (r^*+{r^*}') -4 r^* {r^*}' \cr
\cr
&+&({\Delta t}^2-({r^*}^2+{{r^*}'}^2)) ~ {\ln} |({\Delta t}^2 - (r^*-{r^*}')^2)|\cr
\cr
&-&({\Delta t}^2 - ({r^*}^2+{r^*}'^2) )  ~{\ln}|(-2 r^*_c+r^*+{r^*}')^2-{\Delta t}^2 | \cr
\cr
&-&4 r^*_c (r^*_c + \Delta t) ~{\ln}|-2 r^*_c+r^*+{r^*}'-\Delta t | \cr
\cr
&-&4 r^*_c (r^*_c - \Delta t)~ {\ln} |-2 r^*_c+r^*+{r^*}'+\Delta t|\bigg{)} ~.
\eea
We can simply see that the divergent term (${\ln} \mu$) and the term coming from the regularization ($\gamma$) are not coordinate dependent and can be thrown away. So we will be left with a finite result. 

In addition to the above result we also have to consider the contribution of the zero mode (\ref{zeromode}). The Hadamard function of the zero mode is
\be
\langle 0 | \{X_0(t), X_0(t')\} | 0 \rangle = \frac{-2 \alpha'}{l^2} \pi r^*_c ~\frac{{\cos} \mu\Delta t}{\mu} ~,
\ee
which in the limit $\mu\rightarrow 0$ reduces to 
\be
\langle 0 | \{X_0(t), X_0(t')\} | 0 \rangle = \frac{-2 \alpha'}{l^2} \pi r^*_c ~\frac{1}{\mu} ~.
\ee
This should be added to our previous result for the anti-commutator. But you see the contribution of the zero mode is divergent but coordinate-independent. So it can be thrown away. Therefore our final result for the regularized ${\textrm{AdS}}_3$ Hadamard function which we denote as $g(t,r;t',r')$ is
\bea
\label{AdS-hadamard}
g(t,r;t',r') &=& \frac{ \alpha'}{2 l^2} \bigg{(}12 {r^*_c}^2 -4 r^*_c (r^*+{r^*}') -4 r^* {r^*}' \cr
\cr
&+&({\Delta t}^2-({r^*}^2+{{r^*}'}^2)) ~ {\ln} |({\Delta t}^2 - (r^*-{r^*}')^2)|\cr
\cr
&-&({\Delta t}^2 - ({r^*}^2+{r^*}'^2) )  ~{\ln}|(-2 r^*_c+r^*+{r^*}')^2-{\Delta t}^2 | \cr
\cr
&-&4 r^*_c (r^*_c + \Delta t) ~{\ln}|-2 r^*_c+r^*+{r^*}'-\Delta t | \cr
\cr
&-&4 r^*_c (r^*_c - \Delta t) ~{\ln} |-2 r^*_c+r^*+{r^*}'+\Delta t|\bigg{)} ~.
\eea
We can see that it satisfies the equation of motion for the modes on ${\textrm{AdS}}_3$ background as expected.

\subsection {BTZ Two-Point Functions}
In subsection (\ref{subsec:Mode Expansion in BTZ}) we explicitly wrote the solution to the wave equation on BTZ background. So we are now able to calculate the two-point function of the fields. As has been mentioned in the beginning of the section the commutator of the fields is state-independent and we can calculate it without being worried about the thermal vacuum of the BTZ space. But this is not the case for the Hadamard function and the effect of Hawking radiation should be considered. 

\subsubsection{Green's Function} 
In the previous section we already discussed the details of how to calculate the Green's function. In this section we skip the details and concentrate on the final results. We use the BTZ wave equation solution, eq (\ref{btz-mode}), to write the commutator. One can see that the integrals are very complicated. So in order to be able to obtain the Green's function explicitly we consider the limit where $r_c\rightarrow \infty$. Therefore the commutator gets very simplified and reduces to
\bea
\langle 0|[X(r,t),X(r',t')]|0 \rangle &=& \int_0^\infty dk~ \frac{-4 i \alpha' l^2}{\omega} \frac{  {\sin} \omega(t-t')}{r' r (r_H^2 + k^2 l^4)} \\
~~\cr
&&\bigg{\{}k^2 l^4  ~{\cos} kr^* ~{\cos} k{r^*}' +r r'  ~{\sin}  kr^* ~{\sin} k{r^*}'\cr 
~~\cr
&+&k l^2 r'   ~{\cos} kr^*  ~{\sin} k{r^*}'  +k l^2 r  ~{\cos} k{r^*}'  ~{\sin} kr^* \bigg{\}} ~.\nonumber
\eea    
The integrals are finite as it should be. Therefore the BTZ Green's function  in the limit  $r^*_c \rightarrow 0$, in different regions of figure (\ref{fig:green}), gets the form 
\bea
{\bf I}&:&G(r,r',\Delta t) = 0 ~,\cr
~~\cr
{\bf II}&:& G(r,r',\Delta t) = \frac{2 \pi \alpha' l^2}{r_H^2}   \bigg{[}1-  {\cosh} (\frac{r_H \Delta t}{l^2}) ~{\textrm{sech}} (\frac{r_H r^*}{l^2})  ~{\textrm{sech}} (\frac{r_H {r^*}'}{l^2})\bigg{]} ~,\cr
~~\cr
{\bf III}&:& G(r,r',\Delta t) = 0 ~,\cr
~~\cr
{\bf IV}&:& G(r,r',\Delta t) = \frac{-2 \pi \alpha' l^2}{r_H^2}   \bigg{[}1-  {\cosh} (\frac{r_H \Delta t}{l^2})  ~{\textrm{sech}} (\frac{r_H r^*}{l^2}) ~{\textrm{sech}} (\frac{r_H {r^*}'}{l^2})\bigg{]}~ ,\cr
~~\cr
{\bf V}&:&G(r,r',\Delta t) = 0~ .
\eea
We can see that the Green's function satisfies the equation of motion. Similar to ${\textrm{AdS}}_3$ Green's function in $r^*_c\rightarrow 0$ limit, the Green's function is zero in regions I and V. It is also zero in region III which is outside the light cone. The retarded and advanced Green's functions also appear with opposite signs.

\subsubsection{Hadamard Function} 
To calculate the Hadamard function of the transverse modes in BTZ background we use the same equation as in the ${\textrm{AdS}}_3$ space, (\ref{anti}), where the BTZ wave equation is given in (\ref{inflimit}). Note that in BTZ case the vacuum is thermal and we have 
\be
\langle a^\dagger_k a_{k'} \rangle = \frac{\delta(k-k')}{e^{\beta \omega}-1} ~,
\ee
where $\beta = \frac{1}{T_H}$ and $T_H$ is the Hawking temperature of the black hole, $T_H=\frac{r_H}{2 \pi l^2}$. Therefore
\bea
\langle 0 | \{X(t,r), X(t',r')\} | 0 \rangle &=& \frac{4 \alpha' l^2}{r r'} \int_0^\infty dk~ \frac{1}{k (r_H^2 +k^2 l^4)} ~{\coth} \frac{\beta k}{2}  ~{\cos} k \Delta t\\
\cr
&&(k l^2 ~ {\cos} k r^* + r ~ {\sin} k r^*)(k l^2 ~ {\cos} k {r^*}' + r'  ~{\sin} k {r^*}')~ .\nonumber
\eea

In order to simplify the calculations we would like to obtain the Hadamard function in the limit where $r \rightarrow \infty$. Therefore the integral can be expanded in terms of $r^*$ as
\bea
&&\langle 0 | \{X(t,r), X(t',r')\} | 0 \rangle = \frac{4 \alpha' l^2}{r_H^2} ~\int_0^\infty  dk ~{\cos}  k\Delta t ~{\coth}  \frac{k\beta}{2} \\
\cr
&&\bigg{[}\frac{r_H^2}{9 l^8} k (r_H^2+k^2 l^4) {r^*}^3 {{r^*}'}^3 - \frac{k r_H^2}{90 l^{12}} (4 r_H^4 + 5 r_H^2 l^4 k^2 +  l^8 k^4) ({r^*}^3 {{r^*}'}^5+{r^*}^5 {{r^*}'}^3)+O({r^*}^{10})\bigg{]} ~.\nonumber
\eea
Let's consider the first term in the expansion. The integral over $k$ can be simplified to the form
\be
g(t,r;t',r') = - \frac{4 \alpha' }{9 l^6} ~{r^*}^3 {{r^*}'}^3 ~\int_{-\infty}^{\infty} dk~k(r_H^2+k^2 l^4)~ \frac{{\cos} k\Delta t}{1-e^{\beta k}}~ ,
\ee
where we have used the following relation 
\be
{\coth}  \frac{\beta k}{2} = \frac{1}{1-e^{-\beta k}} - \frac{1}{1-e^{\beta k}}~ .
\ee
The result of the integral is
\be
\label{BTZanti}
g(t,r;t',r') = \frac{\alpha' r_H^4}{6 l^{10}}~ {r^*}^3 {{r^*}'}^3~ \frac{1}{{\sinh}^4 \frac{r_H \Delta t}{2 l^2}}~ ,
\ee
where we have used the fact that $\beta = \frac{1}{T_H} = \frac{2 \pi l^2}{r_H}$. It can be observed that the Hadamard function is periodic in imaginary time, which it should be since the vacuum is thermal.

\section{Two-Point Functions in Vaidya Background}\label{sec:pro} 
In the previous section we mentioned that the Hadamard function of quantum fields which is the expectation value of the anti-commutator depends on the choice of the vacuum. Therefore in a black hole spacetime it encodes the information about the Hawking radiation. In a system with a global time-like killing vector the vacuum is the zero eigenstate of all annihilation operators. The creation and  annihilation operators are associated with positive and negative frequency modes with respect to the time coordinate of the system. In a Vaidya spacetime one can not define one global time-like killing vector for the whole space and therefore the choice of positive frequency modes and the definition of the vacuum is ambiguous. 

To overcome this problem one has to calculate the Bogoliubov coefficients which relate the complex basis of the fields which are solutions to the equation of motion in the vacuum spacetime to the basis in the black hole region. In general it is hard to do this calculation. Therefore people have done it using geometric optics approximation in the late time limit where the spacetime is stationary. This is the whole idea of the late time calculation of Hawking radiation.

In the set-up of this paper we need to address the on-set of Hawking radiation which means to calculate the Hawking radiation in a finite time when the black hole forms. In the AdS/CFT language this calculation  captures the $\frac{1}{N}$ effects. One of the elegant ways of calculating the on-set of Hawking radiation is the work done by Callan, Giddings, Harvey and Strominger (CGHS) \cite{Callan:1992rs} in which they use the trace anomaly method. The set-up of their problem is the formation of a 2-dimensional black hole in linear dilaton background. This method is not applicable to the BTZ black hole formation due to the fact that the 2-dimensional worldsheet problem does not preserve diffeomorphism invariance\footnote{This can be seen from the $r^2$ dependence in front of the third component of the 3-dimensional Vaidya metric which explains the excitations of the string. One can also write this system as a massive scalar field on the 2-dimensional Vaidya background.}. 

In this section we introduce a fairly simple way to calculate the Hadamard function in the Vaidya background which as far as we are aware, has not been used in the literature. The idea is to calculate the Hadamard function of the scalar field modes in the vacuum part of the space and then propagate it to the black hole region using the Green's function in the black hole space. The problem is an initial value problem where the value of the vacuum Hadamard function on the shock wave acts as the initial condition. We expect that this method would give us the information about the on-set of the Hawking radiation. We address this method as the propagation method in the rest of the paper. We will first explain the method more explicitly while applying it to the known example of CGHS \cite{Callan:1992rs}. Then we will use it in the BTZ black hole formation calculation. 


\subsection{Linear Dilaton Black Hole Formation}\label{sub:initial}
In this section we would like to apply the propagation method to the 2-dimensional set-up of formation of the black hole in linear-dilaton background from the collapse of a null shell  \cite{Callan:1992rs}. The linear dilaton vacuum solution is given by
\be
e^{-2\phi} = e^{-2\rho} = - \lambda^2 x^+ x^-~,
\ee
where 
\be
g_{+-} = \frac{-1}{2} e^{2\rho}~,
\ee 
is the metric in the conformal gauge and $(x^+ , x^-)$ are the null coordinates. Using the following change of coordinates the metric reduces to the flat metric:
\be
x^+ = \frac{1}{\lambda}  e^{\lambda y^+}~;~~~~~~~~~ x^- = - \frac{1}{\lambda} e^{- \lambda y^-}~.
\ee
The range of the coordinates $x^+$ and $x^-$ are $(0,\infty)$ and $(-\infty, 0)$, respectively. Therefore $y^-$ and $y^+$ both change over the range $(-\infty,\infty)$.

After the collapse of the null shell (shock wave) at $x^+ = x^+_0$, the black hole solution is given by 
\be
e^{-2\phi} = e^{-2\rho} = a x_0^+ - \lambda^2 x^+ {x^-}'~,
\ee
where $a$ is the amplitude of the shock wave and 
\be
{x^-}' = x^- + \frac{a}{\lambda^2}~,
\ee
is the shift in the null coordinate $x^-$ of the vacuum, due to formation of the horizon. Therefore the coordinate change to flat coordinates for the black hole is given by
\be
x^+ = \frac{1}{\lambda}  e^{\lambda \sigma^+}~;~~~~~~~~~ {x^-}' = - \frac{1}{\lambda} e^{- \lambda \sigma^-}~.
\ee
The relation between the coordinates in the vacuum and black hole regions is 
\bea
y^+ &=& \sigma^+~,\cr
y^- &=& \frac{-1}{\lambda}  ~{\ln}(e^{-\lambda \sigma^-} + \frac{a}{\lambda})~.
\eea
As has been observed in  \cite{Callan:1992rs} the difference between $y^-$ and $\sigma^-$ coordinates is responsible for the nonzero and nontrivial energy-momentum tensor on $\mathcal{I}_R^+$ which gives the Hawking radiation.

The method used in \cite{Callan:1992rs} is based on trace anomaly argument\footnote{We will explain this method in more details in the appendix \ref{app:TA}.} which gives a novel way of calculating a time-dependent Hawking radiation for the process of black hole formation. We would like to redo the calculation using the idea of propagating the two-point function (Hadamard function of the scalar field modes) from the vacuum to the black hole region. To do this we consider the value of the two-point function of the scalar field in the linear dilaton background, as the initial condition and propagate it to the black hole side using the black hole retarded Green's function. 

As the first step we need to obtain the Hadamard function and commutator of the scalar field in the linear dilaton and black hole spacetimes. We first need to solve the equation of motion for the scalar field. The wave equation of motion for the conformal matter is
\be
\partial_+ \partial_- f(x^+,x^-) = 0~.
\ee
Therefore the solution can be decomposed into left moving and right moving parts
\be
f(x^+,x^-) = f_+ (x^+) + f_- (x^-)~.
\ee
Since in this framework the right movers transform the Hawking radiation we can ignore the left moving modes and write $f$ in terms of creation/annihilation operators as
\bea
f_- &=& \int_0^\infty d\omega~ [a_\omega u_\omega + a_\omega^\dagger u_\omega^*]~,\cr
&=& \int_0^\infty d\omega~ [b_\omega v_\omega + b_\omega^\dagger v_\omega^*]~,
\eea
where $u_\omega$ and $v_\omega$ are the wave functions in the vacuum and black hole region, respectively and the creation/annihilation operators satisfy
\be
[a_\omega , a_{\omega '}] =0 ~;~~~~~~ [a^\dagger_\omega , a^\dagger_{\omega '}] =0 ~;~~~~~~ [a_\omega , a^\dagger_{\omega '}] = \delta(\omega-\omega')~. 
\ee
After imposing the boundary conditions at $\mathcal{I}_L^-$ and $\mathcal{I}_R^+$ the wave functions are given by\footnote{Note that for the mode expansion we have followed the notation of \cite{Giddings:1992ff}.} 
\bea
u_\omega &=& \frac{1}{\sqrt{2\omega}} e^{-i\omega y^-}~,\cr
v_\omega &=& \frac{1}{\sqrt{2\omega}} e^{-i\omega \sigma^-}~,
\eea
where they satisfy the normalization conditions
\bea
(u_\omega , u_{\omega '}) &=& (v_\omega , v_{\omega '}) = 2 \pi \delta(\omega-\omega')~,\cr 
(u_\omega , u^*_{\omega '}) &=& (v_\omega , v^*_{\omega '}) =0~.
\eea
Using this mode expansion one can write the Green's function in linear dilaton background as
\bea
G_{\textrm{v}} (y_1^-,y_2^-) &=& -i \langle [f_-(y_1^-) , f_-(y_2^-)]\rangle~,\cr
~\cr
&=& 2 \sum_\omega ~{\textrm{Im}} (u_\omega (y_1^-) u^*_\omega (y_2^-))~,\cr
~\cr
&=& - \frac{\pi}{2} ~{\textrm{sign}} (y_1^- - y_2^-)~.
\eea
Similarly for the black hole background we have
\be
\label{TAgreen}
G_{\textrm{bh}} (\sigma_1^-,\sigma_2^-) = - \frac{\pi}{2} ~{\textrm{sign}} (\sigma_1^- - \sigma_2^-)~.
\ee   
The anticommuator of the modes in the linear dilaton background is given by
\bea
\langle\{f_-(y_1^-) , f_-(y_2^-)\}\rangle &=& 2 \sum_{\omega} ~{\textrm{Re}} (u_\omega (y_1^-) u^*_\omega (y_2^-))~,\cr
~\cr
&=& -\gamma - {\ln} \mu - {\ln} |y_1^- - y_2^-|~.
\eea
Note that the first two terms which come from the regularization are not $y^-$ dependent and can be ignored. Therefore the regularized anticommutator is given by
\be
g_{vacuum} (y^-_1 , y^-_2) =  - {\ln} |y_1^- - y_2^-|~.
\ee
Now we have the necessary tools to calculate the Hadamard function after the formation of black hole using the propagation method.  We would like to propagate each point of the vacuum Hadamard function to the black hole space. In general the propagation formula is  
\be
\label{pro-gen1}
\phi(u,v) = \int du' dv' ~ G(u,v;u',v') J(u',v')~,
\ee
where $(u,v)$ are null coordinates, G is the retarded Green's function of the space and J(u,v) is the source. Let's assume that the shock wave is nonzero at $v=0$. In our set-up there is no source and for $v>0$ we have $\square \phi=0$. But $\phi$ satisfies the initial condition on $v=0$ hypersurface (shock wave) which is $\phi(u,v=0)=\phi_0(u)$. Since $\phi$ is also zero for $v<0$ we can assume that it satisfies the following equation 
\be
\label{initial1}
\square \phi =  \partial_u \phi_0(u) \delta(v)~.
\ee
Comparing this equation with the general Poisson's equation $\square \phi(u,v) = J(u,v)$ we can consider the right hand side of (\ref{initial1}) to act as the source. 

In the set-up of 2-dimensional linear dilaton black hole formation $\phi$ is in fact the black hole Hadamard function $g$ and $\phi_0$ is the vacuum Hadamard function calculated on the shock wave ($x^+=x_0^+$). We can propagate each point from the shock wave using the retarded Green's function, separately. The equation of motion for the retarded Green's function in black hole space is
\be
\partial_{\sigma^+} \partial_{\sigma^-} {\hat G} (\sigma^+,\sigma^-;{\sigma^+}',{\sigma^-}') = \delta(\sigma^+-{\sigma^+}')~\delta(\sigma^--{\sigma^-}')~,
\ee
where 
\be
{\hat G} (\sigma^+,\sigma^-;{\sigma^+}',{\sigma^-}') = - \frac{2}{\pi} G(\sigma^-,{\sigma^-}')~\Theta(\sigma^+-{\sigma^+}')~\Theta(\sigma^--{\sigma^-}')~.
\ee
The propagation formula (\ref{pro-gen1}) for the 2-dimensional set-up of this section reduces to
\be
g_{\textrm{v-bh}} =  \int {d\sigma_1^+}' {d\sigma_1^-}' ~{\hat G}(\sigma_1^+,\sigma_1^-;{\sigma_1^+}',{\sigma_1^-}') ~\partial_{\sigma_1^-} g_0({y_1^-}';{y_2^-}') ~\delta({\sigma_1^+}'-\sigma_0^+)~,
\ee
where $g_{\textrm{v-bh}}$ means the Hadamard function with one point in the vacuum region and the other point in the black hole side. Due to the presence of the $\delta(v)$ function in the definition of the source the integral over ${d\sigma^+}'$ simplifies. We can also see that because of the simple form of the Green's function in black hole space (\ref{TAgreen}) the integrand is a total derivative. Therefore the Hadamard function after the formation of the black hole is given by the vacuum Hadamard function calculated at the end points of the integral. Note that due to the presence of the $\Theta$ function the range of the integration variables ${\sigma_i}'$ is $(-\infty,\sigma_i)$. 

In order to compare our result with the result in CGHS paper we need to calculate the energy-momentum tensor $T_{--}$ which is defined as
\be
\langle 0| T_{--}(\sigma^-) | 0 \rangle = \langle 0| \frac{1}{2} \partial_- f(\sigma^-) \partial_-f(\sigma^-) | 0 \rangle~.
\ee
On the other hand one can see that $T_{--}$ is equal to the regularized $\frac{1}{4} \partial_{\sigma_1^-} \partial_{\sigma_2^-} g(\sigma_1^-,\sigma_2^-)$ where $\sigma_1^- \rightarrow \sigma_2^-$. The derivative of the Hadamard function is 
\bea
\partial_{\sigma_1^-} \partial_{\sigma_2^-} g(\sigma_1^- , \sigma_2^-) &=&  \partial_{\sigma_1^-} \partial_{\sigma_2^-} g(y_1^-(\sigma_1^-) , y_2^-(\sigma_2^-))~, \\
\cr
&=& -  ~\partial_{\sigma_1^-} \partial_{\sigma_2^-} {\ln} \bigg{(}\frac{-1}{\lambda} \bigg{[}{\ln} (\frac{a}{\lambda}+ e^{- \lambda \sigma_1^-})- {\ln} (\frac{a}{\lambda}+ e^{-\lambda \sigma_2^-})\bigg{]}\bigg{)}~, \cr 
\cr
&=& - \lambda^2 ~\frac{\bigg{(}{\ln} (\frac{a}{\lambda}+ e^{- \lambda \sigma_1^-})- {\ln} (\frac{a}{\lambda}+ e^{-\lambda \sigma_2^-})\bigg{)}^{-2}}{(1+\frac{a}{\lambda}e^{\lambda \sigma_1^-}) (1+\frac{a}{\lambda}e^{\lambda \sigma_2^-})}~.\nonumber
\eea
In order to obtain the energy-momentum tensor we have to set $\sigma_1^- = \sigma_2^-$. But it can be easily seen that the Hadamard function is divergent in this limit. To sort out this problem we use point splitting method to eliminate the divergency. Let's assume $\sigma_2^- = \sigma_1^- +\delta$  and then send $\delta$ to zero. Therefore the above result is regularized by subtracting the term $\frac{1}{\delta^2}$. This corresponds to the contribution of the black hole vacuum \cite{Giddings:1992ff}. Therefore we will be able to get the normal ordered energy momentum tensor (in leading order in $\delta$) which is 
\be
\langle T_{--} \rangle = \frac{\lambda^2}{48} \bigg{[}1-\frac{1}{(1+\frac{a}{\lambda} e^{\lambda \sigma^-})^2}\bigg{]}~.
\ee
This is exactly the result obtained in \cite{Callan:1992rs}. Near the horizon which corresponds to $\sigma^-\rightarrow \infty$ the stress tensor reaches the constant value of $\frac{\lambda^2}{48}$ which is the late time Hawking radiation. 

\subsection{BTZ Black Hole Formation}\label{sub:result}

As it was mentioned before we can not use the trace anomaly method to calculate Hawking radiation in our set-up which is  the formation of the BTZ black hole from the collapse of a null shell. One could look at the fluctuations of the string on the other directions of the boundary for example one of the directions on the torus in the background ${\textrm{AdS}}_3 \times T^4 \times S^3$. This calculation is similar to the calculation for a conformal scalar on an ${\textrm{AdS}}_2$ Vaidya background which  has been done in appendix \ref{app:TA}. We will see the instantaneous thermalization of the system after the formation of the black hole which is expected from the discussions of  \cite{Spradlin:1999bn}.  

In this section we would like to repeat the calculation done in the previous section to get the Hadamard function of the BTZ black hole on the boundary.  In order to simplify the problem we consider the limit where $r_c \rightarrow \infty$. So we can compare the result with (\ref{BTZanti}). This corresponds to an infinitely massive quark on the boundary. In this limit the scalar field $X$ which describes the fluctuation of the string (quark on the boundary) behaves like ${r^*}^3$ close to the boundary. This corresponds to the normalizable mode. 

Similar to the previous calculations we start from the ${\textrm{AdS}}_3$ Hadamard function in ${\textrm{AdS}}_3$ background where one of the points is on $v'_1=0$, the shock wave, and the other point has the general value $v'_2$. The analogy is to propagate the $v'_1=0$ point to BTZ boundary. Then we set $v'_2=0$ and propagate the second point to the boundary. We will see that this sorts out the problem of divergency in ${\textrm{AdS}}_3$ Hadamard function when $v'_1=v'_2=0$.   

Therefore the BTZ Green's function (G) and ${\textrm{AdS}}_3$ Hadamard function (g) in the limit $r_c \rightarrow \infty$ get simplified to 
\bea
\label{Green}
G(t,r^*;t',{r^*}') &=& {\hat G}(t,r^*;t',{r^*}') ~{\Theta}(u'-v) {\Theta}(u-u') {\Theta}(v-v')~,\\
\cr
{\hat G}(t,r^*;t',{r^*}') &=& -\frac{2 \pi \alpha' l^2}{r_H^2} \bigg{(}1-{\cosh}(\frac{r_H}{l^2}(t-t')) ~{\textrm{sech}} (\frac{r_H}{l^2} r^*) ~{\textrm{sech}} (\frac{r_H}{l^2} {r^*}')\bigg{)}~, \cr
g(w_1,v_1;w_2,v_2) &=& - \frac{\alpha'}{2 l^2} \bigg{[} -(w_1-v_1) (w_2-v_2)+ \cr
\cr
&& (\omega_1 v_1 +\omega_2 v_2 -\frac{1}{2} (\omega_1+v_1)(\omega_2+v_2)) ~ {\ln} \bigg{|}\frac{(w_1-w_2)(v_1-v_2)}{(w_2-v_1)(w_1-v_2)}\bigg{|}\bigg{]}~,\nonumber
\eea
where $w=t-r^*$ is in fact the coordinate $u$ in ${\textrm{AdS}}_3$ space. The relation between $w$ in ${\textrm{AdS}}_3$ and $u$ in BTZ is 
\be
\label{AdS-BTZ}
w = \frac{2l^2}{r_H} ~{\tanh}(\frac{r_H}{2 l^2} u)~.
\ee
In order to simplify the calculations we will work with the modified two-point functions defined as
\bea
{\hat G}'(t,r^*;t',{r^*}')&=& \frac{-1}{2\pi \alpha' l^2}~r r'{\hat G}(t,r^*;t',{r^*}')~,\cr
\cr
g'(t,r^*;t',{r^*}')&=& r r' g(t,r^*;t',{r^*}')~.
\eea
Note that the modified BTZ Green's function satisfies the following equation of motion
\be
\label{greeneq}
\partial_v\partial_u {\hat G}' + \frac{r_H^2}{2 l^4} ~\frac{1}{{\sinh}^2 (\frac{r_H r^*}{l^2})} {\hat G}' =   {\delta} (u-u') {\delta} (v-v')~.
\ee 
Therefore a general solution to the equation of motion, $\phi(u,v)$, with the source $J(u,v)$  satisfies the equation 
\be
\partial_v\partial_u \phi + \frac{r_H^2}{2 l^4} ~\frac{1}{{\sinh}^2 (\frac{r_H r^*}{l^2})} \phi =  J(u,v)~,
\ee
and can be written as
\be
\phi(u,v) = \int du' dv' ~ G(u,v;u',v') J(u',v')~,
\ee
where $G(u,v;u',v')$ is the Green's function satisying the equation (\ref{greeneq}).  Similar to the discussion in subsection \ref{sub:initial}, one can assume that the source gets the form $J(u,v)=  \partial_u \phi_0 (u,v)~{\delta} (v)$. Therefore $\phi$ can be written in the form
\be
\phi(u,v)=  \int du' G(u,v,u',v'=0)~ \partial_{u'} \phi_0(u')~,
\ee
In the BTZ formation calculation $\phi_0$ is in fact ${\textrm{AdS}}_3$ Hadamard function $\textrm{g}'$ where $v'_1=0$ and ${G}$ is BTZ Green's function, ${\hat{\textrm{G}}}'$. Therefore we have
\bea
g_{{\textrm{AdS}}_3 ; {\textrm{BTZ}}} &=&  \int_0^\infty du'_1~ {\Theta}(u_1'-v_1) {\Theta}(u_1-u_1') \cr
\cr
&&{\hat G}'(u_1,v_1;u'_1,v'_1=0) ~\partial_{u'_1} g' (u'_1,v'_1=0;u'_2,v'_2)~.
\eea
$g_{{\textrm{AdS}}_3 ; {\textrm{BTZ}}}$ is the two-point function where $(u_1,v_1)$ is in BTZ space and $(u_2,v_2)$ is in ${\textrm{AdS}}_3$ space. The ${\Theta}$ functions in the integral change the limits of the integration  to $(v,u)$ which is $(t+r^*,t-r^*)$.  Therefore we can simplify the integral further by changing the integration variable as
\be
u'_1= t_1-x_1 r^*_1~.
\ee
Therefore we get
\be
\label{pro2}
g_{{\textrm{AdS}}_3 ; {\textrm{BTZ}}}= - r^*_1 \int^{1}_{-1} dx_1~{\hat G}'(u_1,v_1;u'_1,v'_1=0) ~\partial_{u'_1} g' (u'_1,v'_1=0;u'_2,v'_2)|_{u'_1= t_1-x_1 r^*_1}~.
\ee
Since we are interested in what happens close to the boundary and we have already sent the boundary to infinity, we expand the integrands around $r^*\rightarrow 0$. Therefore the integral is more simplified  and the final result after integration is
\bea
g_{{\textrm{AdS}}_3 ; {\textrm{BTZ}}}&=& \frac{2}{3 l^2} \rho_1^2 ~\bigg{(}r_H  {\coth} (\frac{r_H t_1}{2 l^2}) \partial_{u_1} g(\omega_1(u_1);u_2,v_2) \cr
\cr
&+& l^2  \partial^2_{u_1} g(\omega_1(u_1);u_2,v_2)\bigg{)}|_{u_1=t_1}~.
\eea
Note that one can take into account the fact that the derivatives of the $g$ with respect to BTZ coordinate $u$ should be changed to the derivatives with respect to ${\textrm{AdS}}_3$ coordinate, $\omega$. We can calculate $g_{{\textrm{AdS}}_3 ; {\textrm{BTZ}}}$ explicitly. Its value at $v'_2=0$ will be the initial data for the propagation of the second point to the boundary
\be
g^0_{{\textrm{AdS}}_3 ; {\textrm{BTZ}}} =  \frac{2 \alpha' r_H^4}{3 l} ~\rho_1^2~\frac{\omega_2^2}{(-2 l^2 (-1+ {\cosh}  (\frac{r_H t_1}{l^2}))+r_H \omega_2 ~{\sinh} (\frac{r_H t_1}{l^2}) )^2}~,
\ee  
where we have used (\ref{AdS-hadamard}).
Now we can use the same integral (\ref{pro2}) and propagate the second point to the boundary.  The final result after propagation is
\be
g'_{{\textrm{BTZ}}} =  \frac{\alpha'  r_H^4}{6 l^6}~ {r^*_1}^2 {r^*_2}^2 ~\frac{1}{{\sinh}^4 (\frac{r_H(t_1-t_2)}{2 l^2})}~. 
\ee
Therefore the BTZ Hadamard function is
\be
g_{{\textrm{BTZ}}} =  \frac{\alpha'  r_H^4}{6 l^{10}}~ {r^*_1}^3 {r^*_2}^3 ~\frac{1}{{\sinh}^4 (\frac{r_H(t_1-t_2)}{2 l^2})}~. 
\ee
One can see that this result is exactly the same as the result we got before for BTZ Hadamard function  (\ref{BTZanti}). As we can see here the BTZ Hadamard function is periodic in imaginary time with the period $T_H^{-1}$. Therefore it represents  a thermal state at temperature $T_H$. This implies instantaneous thermalization on the boundary of the 3-dimensional Vaidya spacetime for non-local operators like the Hadamard function. 

One could also assume a detector at fixed r which will immediately after the shock wave will feel the thermal bath.  

This result is also similar to the result obtained in \cite{Spradlin:1999bn} for the correlation function of the boundary operator for ${\textrm{AdS}}_2$ spacetime, using AdS/CFT duality. In the small separation limit where $\Delta t  \ll 1$ it behaves like $\frac{1}{{\Delta t}^4}$ and for large separation  $\Delta t \gg 1$ it decays exponentially $e^{-4 \Delta t}$. 

\newpage
\noindent{\bf{\large Acknowledgment}}

~

We would like to thank B. Chakraborty, S. Detournay, S. Hartnoll, V. E. Hubeny, N. Iqbal, A. Lawrence, S. Minwalla, M. Rangamani, M. Shigemori and A. Strominger for stimulating discussions. We are grateful to V. E. Hubeny, M. Rangamani and A. Strominger for comments on the manuscript.  M. H. would also thank Harvard University for warm hospitality. This work was supported in part by the DOE under grant No. DE-FG02-92ER40706.

~~~~ 

~~~

\appendix
\noindent{\large{\bf{Appendix}}}
\section{${\textrm{AdS}}_2$ Black Hole Formation}\label{app:TA}

In this appendix we show how the calculation for trace anomaly on ${\textrm{AdS}}_2$ black hole formation predicts the rapid thermalization. This can be seen as the fluctuations of the string in the other directions of the space ${\textrm{AdS}}_3\times {\bf M}$ where ${\bf M}$ can be $T^4\times S^3$. For example the transverse direction can be chosen to be along one of the torus coordinates. This problem is similar to solving a scalar field equation on ${\textrm{AdS}}_2$ Vaidya background. The system satisfies all the necessary conditions to use trace anomaly method; there is a conformal scalar and the set-up preserves diffeomorphism invariance.  We will do this calculation using both trace anomaly method and propagation of the Hadamard function of the scalar field from the vacuum to the black hole space.  

The energy-momentum tensor of the 2-dim theory is
\be
T_{\mu\nu} = \frac{1}{2\pi \alpha'} \bigg{(}\frac{-1}{2} g_{\mu\nu} g^{\rho\sigma} G_{IJ} \partial_\rho X^I \partial_\sigma X^J + G_{IJ} \partial_\mu X^I \partial_\nu X^J\bigg{)}~,
\ee
where we have used the action (\ref{action}). Therefore we explicitly get
\bea
T_{rt} &=& \frac{1}{2\pi \alpha'} \frac{r^2}{l^2} (\partial_r X \partial_t X)~,\cr
\cr
T_{tt} &=& \frac{1}{4\pi \alpha'} \frac{r^2}{l^2} (\partial_t X \partial_t X - g_{tt}^2 \partial_r X \partial_r X)~,\cr
\cr
T_{rr} &=& \frac{1}{4\pi \alpha'} \frac{r^2}{l^2} (\partial_r X \partial_r X - g_{rr}^2 \partial_t X \partial_t X)~.
\eea 
One can also write the energy-momentum tensor in null coordinates which would be
\bea
\label{def-em-tensor}
T_{uv} &=& 0~,\cr
\cr
T_{vv} &=& \frac{1}{2\pi \alpha'} \frac{r^2}{l^2} (\partial_v X \partial_v X)~,\cr
\cr
T_{uu} &=& \frac{1}{2\pi \alpha'} \frac{r^2}{l^2} (\partial_u X \partial_u X)~.
\eea
Using these coordinates one can see that the shock wave in the Penrose diagram in figure (\ref{fig:penrose})\footnote{Note that the Penrose diagram for the ${\textrm{AdS}}_2$ black hole formation is similar to the BTZ black hole formation except that there is no singularity in the ${\textrm{AdS}}_2$ case and the black hole is explained by the coordinate which covers only a part of the space \cite{Spradlin:1999bn}.} travelling along $u$ direction with the amplitude $a$ corresponds to 
\be
T_{vv} = a \delta(v)~.
\ee 
Using the argument of trace anomally the expectation value of the energy-momentum tensor is given by
\be
\label{trace anomaly}
\langle T\rangle = \frac{c R}{24 \pi}~.
\ee
where $c$ is the central charge which is one for the conformal matter (scalar field) $X$ and $R$ is the ricci scalar which is the constant $\frac{-2}{l^2}$ for the ${\textrm{AdS}}_2$ and asymptotically ${\textrm{AdS}}_2$ black hole spacetime. We work in the conformal gauge where we can write the metric as
\be
g_{vu} = \frac{-1}{2} e^{2 \rho} ~; ~~~~~g_{vv} = g_{uu} =0~.
\ee  
Therefore using (\ref{trace anomaly}) the conservation of the enrgy-momentum tensor implies 
\bea
\label{emtensor}
\langle T_{vu}\rangle &=& - \frac{1}{12 \pi} \partial_v \partial_u \rho ~,\cr
\cr
\langle T_{vv}\rangle &=& - \frac{1}{12 \pi} ( \partial_v \rho \partial_v \rho - \partial^2_v \rho + t_v(v))~,\\
\cr
\langle T_{uu}\rangle &=& - \frac{1}{12 \pi} ( \partial_u \rho \partial_u \rho - \partial^2_u \rho + t_u(u))~.\nonumber
\eea
where $t_v$ and $t_u$ are integration constants which should be fixed by boundary conditions. Note that for the empty ${\textrm{AdS}}_2$ space, where there is no nonzero energy flux or radiation, both $T_{uu}$ and $T_{vv}$ are zero while $T_{uv}$ is nonzero and produces the ${\textrm{AdS}}_2$ curvature.  

\subsubsection{Scalar Field on ${\textrm{AdS}}_2$ Vaidya Background}
The first step to calculate the energy-momentum tensor is to find the right coordinates for the Vaidya background which keeps the metric continuous across the shock wave (metric itself, not necessarily its derivatives). We can use the calculation in section \ref{sec:set-up} which led to (\ref{u-relation}). Assuming the metric has the general form  
\be
ds^2 = - e^{2\rho} du dv~,
\ee
in the two spaces of the problem we get 
\bea
ds^2 &=& \frac{- r_H^2}{l^2} ~\frac{1}{{\sinh}^2(\frac{r_H}{2 l^2} (v-u))}~dv du~;~~~~~~v<0\\
\cr
ds^2 &=& \frac{-4 l^2}{\bigg{(}v  ~{\cosh}(\frac{r_H}{2 l^2} u) -\frac{2 l^2}{r_H} ~{\sinh}(\frac{r_H}{2 l^2} u)\bigg{)}^2} ~dv du~;~~~~~~v>0\nonumber
\eea 
where u is the null coordinate in ${\textrm{AdS}}_2$ black hole space.  One can see that the metric is continuous across the shock wave $v=0$, as it was expected. Therefore we get for $v>0$
\be
(\partial_v \rho)^2 - \partial_v^2 \rho = (\partial_{u} \rho)^2 - \partial_{u}^2 \rho = \frac{r_H^2}{4 l^4}~,
\ee
and for $v<0$
\bea
(\partial_v \rho)^2 - \partial_v^2 \rho &=& 0~, \cr
\cr
(\partial_{u} \rho)^2 - \partial_{u}^2 \rho &=& \frac{r_H^2}{4 l^4}~.
\eea
At this stage we have to impose the boundary conditions to fix the integration constants $t_v$ and $t_u$.  We know that in the vacuum region the $T_{uu}$ and $T_{vv}$ are zero. This implies that $t_v = 0$ and $t_u = -\frac{r_H^2}{4 l^4}$.  Therefore in the black hole region we get
\bea
\label{Tvv}
\langle T_{vv}\rangle &=& - \frac{r_H^2}{48 \pi l^4}~,\cr
\cr
\langle T_{uu}\rangle &=& 0~.
\eea
This shows that the energy-momentum tensor component $T_{vv}$  is zero for $v<0$ but nonzero and constant after the shock wave. This happens due to the collapse of the null shell. Note that in contrast to linear dilaton black hole formation in \cite{Callan:1992rs}, the energy-momentum observed in the black hole region changes from zero to a constant value instantly. Therefore one can conclude that the observer on the boundary instantly feels the thermal bath of the Hawking radiation due to the formation of the black hole. This is similar to the argument which has previously been given in \cite{Spradlin:1999bn}.  

We can also see that the curvature, $R=8 e^{-2\rho} \partial_v \partial_\omega \rho$ is constant and the same for all values of $v$ and is given by $\frac{-2}{l^2}$. 

In the next subsection we will compare the trace anomaly method result with the propagation of the Hadamard function. 

\subsubsection{${\textrm{AdS}}_2$ Vaidya Spacetime Two-Point Functions}
In order to use the propagation idea we first need to calculate the two-point functions in ${\textrm{AdS}}_2$ and ${\textrm{AdS}}_2$ black hole spaces. Since we just have the $(t,r)$ coordinates of ${\textrm{AdS}}_3$ and BTZ spaces the equation of motion simplifies and the wave equation is
\be
u_{\omega}(t,r^*) = \sqrt{\frac{2\alpha'}{\omega}} ~{\sin} \omega(r^*-r^*_c)~ e^{-i \omega t}~.
\ee 
The form of the wave equation is the same for both spaces and the only difference is in the definition of $r^*$. 
So the Green's function (G) for both spaces is given by
\bea
G(t,r^*;t',{r^*}') &=& \frac{\alpha' \pi}{2} \bigg{[} - {\textrm{sign}}(\Delta t+r^*-{r^*}') - {\textrm{sign}}(\Delta t-r^*+{r^*}')\\
\cr
&+&{\textrm{sign}}(\Delta t+r^*+{r^*}'-2r^*_c) + {\textrm{sign}}(\Delta t-r^*-{r^*}'+2r^*_c)\bigg{]}~,\nonumber
\eea
where $\Delta t = t-t'$. One can see that the retarded Green's function which is
\be
{\hat G} (v,u;v',u') = - \frac{1}{\pi \alpha'} ~G(v,u;v',u') ~\Theta(u-u')~ \Theta(v-v')~,
\ee
satisfies the equation of motion with the $\delta$ function source
\be
\partial_v \partial_u  {\hat G} =  \delta(u-u') ~\delta(v-v')~.
\ee
The Green's function in different regions of the light cone diagram, figure \ref{fig:green}, are given by
\bea
\label{green-2}
{\textrm{I}}~&:&~~~~G(t,t';r^*,{r^*}') = 0~,\cr
\cr
{\textrm{II}}~&:&~~~~G(t,t';r^*,{r^*}') = \alpha' \pi~,\cr
\cr
{\textrm{III}}~&:&~~~~G(t,t';r^*,{r^*}') = 0~,\\
\cr
{\textrm{IV}}~&:&~~~~G(t,t';r^*,{r^*}') = -\alpha' \pi~,\cr
\cr
{\textrm{V}}~&:&~~~~G(t,t';r^*,{r^*}') = 0~.\nonumber
\eea
The other two-point function is the Hadamard function of the scalar field in the ${\textrm{AdS}}_2$  and ${\textrm{AdS}}_2$ black hole space (similar to the Green's function the difference is only in the definition of $r^*$) which is given by
\bea
\label{anti-2}
g(t_1,{r_1^*};t_2,{r_2^*}) &=& \alpha'  \bigg{[} - {\textrm{ln}} |t_1-t_2+{r_1^*}-{r_2^*}| - {\textrm{ln}} |t_1-t_2-{r_1^*}+{r_2^*}|\\
\cr
&+&{\textrm{ln}} |t_1-t_2+{r_1^*}+{r_2^*}-2r^*_c| + {\textrm{ln}} |t_1-t_2-{r_1^*}-{r_2^*}+2r^*_c|\bigg{]}~,\nonumber
\eea
which also satisfies the equation of motion for each variable.  

Now we have necessary tools to calculate the Hadamard function of the scalar field in the ${\textrm{AdS}}_2$ Vaydia space. We would like to propagate each point of the ${\textrm{AdS}}_2$ Hadamard function to the ${\textrm{AdS}}_2$ black hole space. We will follow the same analogy as section \ref{sec:pro}. The initial condition in the propagation formulas (\ref{pro-gen1}) and (\ref{initial1}) is given by $AdS_2$ Hadamard function calculated on the shock wave ($v=0$).  One can see that the ${\textrm{AdS}}_2$ Hadamard function (\ref{anti-2}) is divergent when one sets $v_1=v_2=0$. To avoid this divergency we propagate each point individually, which means we first set $v'_1=0$ and propagate the first point and then we set $v'_2=0$ and propagate the second point. These words translated into formulas give
\bea
g_{\textrm{v-bh}} &=& - \frac{1}{\pi \alpha'} \int_0^\infty du'_1 ~ \Theta(u_1-u'_1) \Theta(u'_1-v_1+2 r^*_c)\cr
\cr
&&  G(v_1,u_1;v'_1=0,u'_1) ~\partial_{u'_1} g(v'_1=0,\omega'_1,v'_2,\omega'_2)~.
\eea 
Using the relation for the Green's function (\ref{green-2}) and considering the range of $u'_1$ imposed by the $\Theta$ functions the above propagation relation reduces to
\be
\label{pro-2}
g_{\textrm{v-bh}} =  \int_{v_1-2 r^*_c}^{u_1} du'_1~ \partial_{u'_1} g(v'_1=0,\omega'_1,v'_2,\omega'_2)~,
\ee
where the relation between the null coordinate $u$ in ${\textrm{AdS}}_2$ black hole and the null coordinate $\omega$ of ${\textrm{AdS}}_2$ are given in (\ref{u-relation}). One can see that the integrand is the total derivative and the result is the ${\textrm{AdS}}_2$ Hadamard function calculated in the end points of the integral. Note that the terms in g (\ref{anti-2}) which don't depend on $u'_1$ cancel which means that the divergent term, ${\textrm{ln}} |v'_1-v'_2|$, will be removed. Therefore we have
\bea
g_{\textrm{v-bh}} &=& - \alpha' \bigg{[} {\textrm{ln}} |\omega_1(u_1) - \omega'_2| - {\textrm{ln}} |\omega_1(v_1-2 r^*_c) - \omega'_2|\cr
\cr
&-& {\textrm{ln}} |\omega_1(u_1) - v'_2+2 r^*_c| + {\textrm{ln}} |\omega_1(v_1-2 r^*_c) - v'_2+2 r^*_c| \bigg{]}~.
\eea
Now we can propagate the second point. So we set $v'_2=0$ in $g_{\textrm{v-bh}}$ and use the propagation formula (\ref{pro-2}) with the integration variable $u'_2$. One can see that only the terms which depend on $\omega'_2$ will be left and we get
\bea
g_{\textrm{bh}} &=&   \alpha' \bigg{[} {\textrm{ln}} |\omega_1(u_1) - \omega_2(u_2)| - {\textrm{ln}} |\omega_1(v_1-2 r^*_c) - \omega_2(u_2)|\cr
\cr
&-&  {\textrm{ln}} |\omega_1(u_1) - \omega_2(v_2-2 r^*_c)| + {\textrm{ln}} |\omega_1(v_1-2 r^*_c) - \omega_2(v_2- 2 r^*_c)|\bigg{]}~.
\eea
We are interested in the relation for the energy-momentum tensor. The energy-momentum tensor should be given by the regularized derivatives of the Hadamard function. In other words we have
\be
\langle T_{u_1 u_2}\rangle = \frac{1}{4\pi \alpha'} \partial_{u_1} \partial_{u_2} g|_{\textrm{regularized}}~,
\ee
and the same for $T_{v_1 v_2}$. The derivatives of the Hadamard function are 
\bea
\partial_{u_1} \partial_{u_2} g&=&  \frac{r_H^2 \alpha'}{4 l^4}~ \frac{1}{{\textrm{sinh}}^2\frac{r_H}{2 l^2}(u_1-u_2)}~,\cr
\cr
\partial_{v_1} \partial_{v_2} g &=&  \frac{r_H^2 \alpha'}{4 l^4}~ \frac{1}{{\textrm{sinh}}^2\frac{r_H}{2 l^2}(v_1-v_2)}~.
\eea
To regularize the above result we use the point-splitting method in which we assume $u_2 = u_1+\epsilon$ where $\epsilon \rightarrow 0$. The divergent part is $\frac{-\alpha'}{\epsilon^2}$ which we can throw away. Therefore we are left with
\be
\langle T_{vv}\rangle = \langle T_{uu}\rangle = - \frac{r_H^2}{48 \pi l^4}~.
\ee
This is similar to what we got in the previous section using trace anomaly method (\ref{Tvv}).


\end{document}